\documentclass[superscriptaddress,twocolumn,aps,pre]{revtex4}
\usepackage{graphicx}% Include figure files

\begin{document}
\title{{\bf Theory of the Three-Group Evolutionary Minority Game}}
\author{Kan Chen}
\affiliation{Department of Computational Science, Faculty of Science, National University of Singapore, Singapore 117543}
\author{Bing-Hong Wang}
\affiliation{Department of Modern Physics, University of Science and Technology of China, Hefei, Anhui, 230026, China}
\author{Baosheng Yuan}
\affiliation{Department of Computational Science, Faculty of Science, National University of Singapore, Singapore 117543}

\date{\today }

\begin{abstract}
Based on the adiabatic theory for the evolutionary minority game (EMG) that we proposed earlier\cite{chen}, we perform a detail analysis of the EMG limited to three groups of agents. We derive a formula for the critical point of the transition from segregation (into opposing groups) to clustering (towards cautious behaviors). Particular to the three-group EMG,  the strategy switching in the ``extreme" group does not occur at every losing step and is strongly intermittent. This leads to an correction to the critical value of the number of agents at the transition, $N_c$. Our expression for $N_c$ is in agreement with the results obtained from our  numerical simulations.

\end{abstract}

\maketitle


\section{Introduction}

There have been growing interests in study of complex adaptive systems using agent-based modeling. Complex adaptive systems are ubiquitous in social, economics, and biological sciences; they consist of agents using adaptive strategies to compete for limited resources. As changes in the global environment are induced by the agents themselves, it is important to study dynamics of such systems. Even though  agent-based models are simple, the outcomes may not be at all obvious because the agents typically use adaptive rather than optimizing strategies.

The Minority Game, proposed by Challet and Zhang\cite{challet1,challet2}, is a prototypical agent-based model that can be analyzed using the tools of statistical mechanics. The game captures some essential features of complex adaptive systems in which the agents with limited information and rationality compete for limited resources. The key question in the study of agent-based models is, how evolution changes the behaviors of the agents. In the context of a simple evolutionary minority game (EMG), Johnson and coworkers found that the agents universally self-segregate into two opposing extreme groups.\cite{johnson} Hod and Nakar, on the other hand, claimed that a clustering of cautious agents emerges  in a ``tough environment'' where the penalty for losing is greater than the reward for winning.\cite{hod1} 
We have derived a general formalism to understand the dynamical mechanism for the transition from segregation to clustering.\cite{chen} Our theory is based on an adiabatic approximation, in which short-time fluctuations are integrated out to obtain a steady state population distribution. We found that the effective rate of evolution plays an important role in determining the resulting steady-state population distribution. Frequent strategy switching leads to large market inefficiency that favors clustering of cautious agents. The theory is illustrated with a detail statistical mechanical analysis of the EMG limited to three groups of agents: two opposing groups and one cautious group; it agrees very well with the numerical simulations of the original EMG, but deviates from the numerical results of the three-group EMG. In this paper we perform a further analysis of the three-group EMG. We show that with an intermittency correction, which is particular to the three-group EMG, the numerical results can be explained.

\section{Phenomenology of the Evolutionary Minority Game}

We first briefly describe the EMG model. There are N agents. At each round they choose to enter room 0 or room 1. At the end of each round the agents in the room with fewer agents (in the minority) win a point; while the agents in the room with more agents lose a point. The winning room numbers (0 or 1) are recorded. The agents' decisions
are based on the same record (a bit-string of length $m$) of the most recent winning room numbers. Given the current $m$-bit string, the common basic strategy is simply to choose the winning room number after the most recent pattern of the same $m$-bit string in the historical record. To use the basic strategy is thus to follow the trend. In the EMG each agent is assigned a probability $p$: he will adopt the basic strategy with probability $p$ and adopt the opposite of the basic strategy with probability $1-p$. The agents with $p=0$ or $1$ are ``extreme'' players, while the agents with $p=1/2$ are cautious players. The game and its outcomes evolve as the less successful agents, defined as the ones with the accumulated wealth less than $d$ ($d<0$), change their $p$ values. In the original EMG model, the new p value is chosen randomly in the interval of width $\Delta p$ centered around its original $p$ value. His wealth is reset to zero and the game continues. 

Johnson and coworkers showed that the agents self-segregate into two opposing extreme groups with $p\sim 0$ and $p\sim 1$.\cite{johnson,lo1,lo2} Thus, in order to succeed in a competitive society the agent must take extreme positions. This behavior can be explained by the market impact of the agents' own actions which largely penalizes the cautious agents.\cite{lo2} However, Hod and Nakar later found that the above conclusion is only robust when the reward-to-fine ratio $R \ge 1$. When $R<1$ there is tendency for the agents to cluster towards cautious behaviors and the distribution of the $p$ value, $P(p)$, may evolve to an inverted-U shape with the peak at the middle. We show that the transition from segregation to clustering in fact depends on all three parameters, $N$, $R$, and $d$. We derived a general
expression\cite{chen} for the critical value $N_c =\left[\frac{|d|}{A(1-R)}\right]^2$,
where $A$ is a constant of the order one. This was verified by our extensive simulations of the EMG for a wide range of the parameter values.\cite{chen} When $R\rightarrow 1$ the clustering only occurs for either very large $N$ or very small $|d|$. At $R=1$ the clustering disappears and the segregation to extreme behaviors becomes robust.

\section{Description of the General Adiabatic Theory for the EMG}

We now briefly describe our theory, illustrated with the three-group EMG model, in which $p$ takes only one of the three possible values $0,1/2$, and $1$. The agents in group 0 (with $p=0$) make the opposite decision to the agents in group 1 (with $p=1$). We denote the group with $p=1/2$ group m; this is the group of cautious agents. 

We begin by evaluating the average wealth reduction for the agents in each of the three groups, given the number of agents  $N_0, N_m, N_1$ in group 0, m, and 1 respectively. By comparing the average wealth reduction in the extreme groups and in the cautious group, we can determine the transition from clustering to segregation. Let $n$ be the number of agents in group m making the same decision (decision A) as those in group 0. When $N_m \gg 1$, the distribution of $n$ can be approximated by a Gaussian distribution $P(n) = \frac{1}{\sqrt{2\pi}\sigma_m}\exp(-(n-N_m/2)^2/(2\sigma_m^2)),$
where $\sigma_m = \sqrt{N_m}/2$. By averaging over $n$, we obtain the average change of wealth for the agents in group 0,
\begin{equation}
\Delta w_0  = -\frac{1-R}{2} + \frac{1+R}{2}\mathrm{erf}\left(\frac{N_d}{2\sqrt{2}\sigma_m}\right),
\end{equation}
where $N_d = N_1 - N_0$;  $\mathrm{erf}(x)$ is the error function 
$\frac{2}{\sqrt{\pi}}\int^x_0e^{-t^2}dt$. Similarly the average change of wealth for the agents in group 1 is given by,
\begin{equation}
\Delta w_1  = -\frac{1-R}{2} - \frac{1+R}{2}\mathrm{erf}\left(\frac{N_d}{2\sqrt{2}\sigma_m}\right).
\end{equation}
The average change of wealth of the agents in the extreme groups (group 0 and 1) is given by 
$\Delta w_e= (N_0  \Delta w_0 + N_1 \Delta w_1)/(N_0 + N_1)$, or
\begin{equation}
\Delta w_e =-\frac{1-R}{2} - \frac{1+R}{2}\frac{N_d}{N_0 + N_1}\mathrm{erf}\left(\frac{N_d}{2\sqrt{2}\sigma_m}\right).
\end{equation}

The second term in $\Delta w_e$ is due to the fluctuations of $N_d$; it is always negative. The larger
the fluctuation in $N_d$, the less efficient the market becomes. This term can be interpreted as the cost due to market inefficiency. Large market inefficiency on average penalizes the players taking ``extreme" positions. 

The average wealth reduction for the agents in group m can be evaluated similarly by averaging over $n$, 
\begin{equation}
\Delta w_m = -(1-R)/2 -\frac{1+R}{\sqrt{2\pi N_m}} \exp(-N_d^2/(2N_m)).
\end{equation}
The first term in $\Delta w_m$ is the same as that in $\Delta w_e$. The second term can be interpreted as the market impact.\cite{lo2}  A large market impact (self-interaction) penalizes the cautious players; their own decisions increase their chances of being in the majority and hence increase their chances of losing. The relative magnitudes of the second terms in $\Delta w_e$ and $\Delta w_m$ determine whether clustering or segregation dominates.

To further evaluate $\Delta w_e$ and $\Delta w_m$, we need to average over the distribution of $N_d$. Let $\delta N$ denote the change in $N_d$ in one time step.  As argued in Ref.~\cite{chen}, the steady state probability distribution $Q(N_d)$ for $N_d$ should satisfy
\begin{eqnarray}
Q(N_d) &=& W_{-}(N_d +\delta N)Q(N_d + \delta N) \nonumber \\
	& &+W_{+}(N_d - \delta N))Q(N_d - \delta N),
\end{eqnarray}
where $W_{\pm} = \frac{1}{2}[1 \mp \mathrm{erf}(N_d/(2\sqrt{2}\sigma_m)]$. This equation can be solved to give the distribution $Q(N_d)$.  
After averaging $N_d$ over $Q(N_d)$, we obtain
\begin{equation}
\Delta w_e = -\frac{1-R}{2}-\frac{(1+R)}{2}\frac{\delta N}{2(N_0 + N_1)}.
\end{equation}
$\Delta w_m$, on the other hand, is given by 
\begin{equation}
\Delta w_m  \sim  -\frac{1-R}{2}-\frac{1+R}{\sqrt{2\pi}}\frac{1}{\sqrt{N_m +\sigma_d^2}}.
\end{equation}
Here $\sigma_d = \sqrt{\frac{\sqrt{2\pi}}{2}}\sqrt{\sigma_m \delta N}$.  
At the critical point, $N_0=N_1=N_m = N/3$, and $\Delta w_e = \Delta w_m$. It is easy to verify that this occurs when 
\begin{equation}
\delta N \sim \sqrt{N}.
\end{equation}  
Note that $\delta N$ is the number of extreme agents switching their strategies per time step. On average
 $\delta N = 2 N_0/(|d|/((1-R)/2))=N_0(1-R)/|d|$, thus the crossover value for $N$ is $N_c = d^2/[A(1-R)]^2$.
As shown in Ref.~\cite{chen} the derivation can be generalized to a general EMG involving more than three groups of agents.

\section{Gap Distribution and Intermittency Correction}

While the expression $N_c = d^2/[A(1-R)]^2$ describes very well the transition in the original EMG with a continuous distribution of $p$, it does not describe the $R$ dependence of $N_c$ correctly for the transition in the three-group EMG. This is due to the following particular feature of the three-group EMG. The strategy switching of the agents in the ``extreme" groups does not occur at every losing step and is rather intermittent. A loss at the current round, for example, will not make the agents in the extreme groups to switch strategy if they have won in the previous two rounds. This gives rise to a gap $\Delta$ between the lowest value of the wealth in an
extreme group and the strategy-switching threshold at $d$, as schematically illustrated in Figure 1. 

\begin{figure}
\includegraphics*[width=8.5cm]{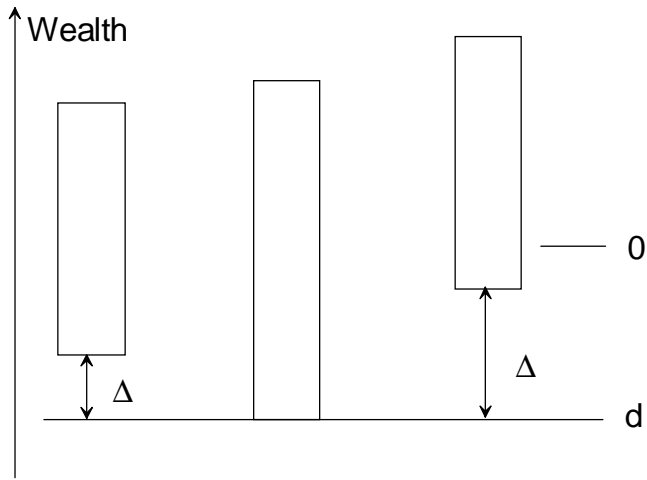}
\caption{Schematic illustration of the wealth distributions of group 0, m, and 1. 
There are gaps in the distributions for the extreme group 0 and 1}
\end{figure}

When  the wealth distribution of the losing extreme group has $\Delta > 1$, $N_d$ does not change, because no switching occurs in or out of the extreme groups. Only the cases with $\Delta < 1$ affect the distribution of $N_d$. Thus
we need to use the effective number of extreme agents switching their strategies per time step, averaged only over the cases with $\Delta < 1$. This effective rate of switching is given by
 $\delta \tilde{N} = \delta N / z = N_0(1-R)/(z|d|)$, where $z$ is the probability that $\Delta < 1$. The effect of intermittency of the strategy switching can be taken into account by using this effective $\delta \tilde{N}$.

To obtain the intermittency correction to the expression of $N_c$, we need to first calculate the probability $z$. Let $P(\Delta)$ be the probability distribution of $\Delta$.  Since there are always new agents with wealth $0$ coming from group m at each step, $\Delta$ cannot be greater than $|d|$. Thus $P(|d|)=0$. Since there are roughly equal numbers of winning and losing steps, $P(\Delta)$ should satisfy
\begin{equation}
P(\Delta) = \frac{1}{2}P(\Delta + 1) + \frac{1}{2}P(\Delta - R).
\end{equation}
For $|d| \gg 1$ we can approximate the above equation using the following differential equation (this is adequate as $P(\Delta)$ is rather smoothly varying as a function of $\Delta$).
\begin{equation}
P^{''}(\Delta) + \lambda P'(\Delta) = 0,
\end{equation}
where $\lambda = \frac{2(1-R)}{1+R^2}$. Given the condition $P(|d|)=0$, the solution of the equation is
\begin{equation}
P(\Delta) = B(e^{-\lambda \Delta}-e^{-\lambda |d|}).
\end{equation}

\begin{figure}
\includegraphics*[width=8.5cm]{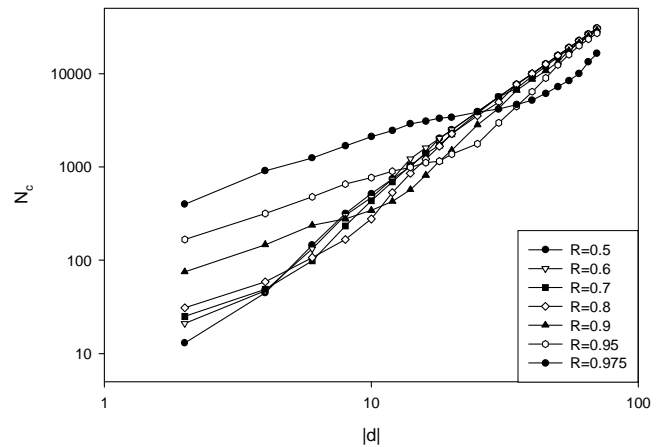}
\caption{The critical value $N$ vs $|d_c|$ for $R=0.6,0.7,0.8,0.9,0.95$, and $0.975$}
\end{figure}

$B$ is determined using the normalization condition $\int^{|d|}_0 P(\Delta)d\Delta = 1$:
$$B=\frac{\lambda}{1-e^{-\lambda |d|} - \lambda |d| e^{-\lambda |d|}}.$$
Now $z$ is given by 
\begin{equation}
z = \int^{1}_{0}P(\Delta)d \Delta = \frac{1-e^{-\lambda} - \lambda e^{-\lambda |d|}}
	{1-e^{-\lambda |d|} - \lambda |d| e^{-\lambda |d|}}
\end{equation}
The critical value $N_c$ is determined from the relation $\delta \tilde{N}_c \sim \sqrt{N_c}$; this gives rise to
\begin{equation}
N_c \sim  (zd)^2/(1-R)^2.
\end{equation}

 The numerical values of $N_c$ vs $|d|$ from our extensive simulations are plotted in Figure 2. One notable feature is that $N_c$ is proportional to $d^2$, but is independent of $R$ for sufficiently large values of $|d|$. This can be explained using the above expression of $N_c$. For $|d|(1-R) \gg 1$ and $1-R \ll 1$, $z$ can be
approximated by $z \sim \lambda \sim 1-R$. This gives rise to $N_c \sim d^2$ for $|d| \gg 1/(1-R)$. From the figure, we can see the crossover to the $N_c \sim d^2$ behavior occurs roughly at $d = 2/(1-R)$, in agreement with the theory. This unique feature of
 the three-group EMG model is completely due to the existence of a gap in the wealth distribution of the extreme groups. For the original EMG model, there is no such gap, and $N_c \sim d^2/(1-R)^2$ holds very well.

\section{Conclusion}

In conclusion, we have derived an intermittency correction to the critical value $N_c$ for the three-group EMG.  Our estimates of $N_c$ agree well with the numerical results. We have identified the key difference between the three-group EMG and the original EMG with a continuous distribution of the p values: the existence of a gap in the wealth distribution of the extreme groups in the three-group EMG. This key difference leads to a qualitative correction to the expression for $N_c$. The general framework of the adiabatic approximation, however, is equally valid for studying the transition in both the original EMG and the three-group EMG. 

\section*{Acknowledgements}

This work was supported by the National University of Singapore research grant R-151-000-028-112. BHW also acknowledges the support by the Special Funds for National Basic Research Major Project ``Nonlinear Science", by the National Natural Science Foundation of China (Nos.19932020,
19974039 and 70271070), and by the China-Canada University Industry
Partnership Program (CCUIPP-NSFC Grant No. 70142005).


\begin{thebibliography}{99}

\bibitem{chen} K. Chen, B.-H. Wang, and B. Yuan, ``Adiabatic Theory for the Population Distribution in the Evolutionary Minority Game'', to be published in Phys. Rev. E (Rapid Communication).

\bibitem{challet1}
D. Challet and Y.-C. Zhang, Physica A {\bf 246}, 407 (1997).

\bibitem{challet2}
D. Challet and Y.-C. Zhang, Physica A {\bf 256}, 514 (1998).

\bibitem{johnson} N. F. Johnson, P. M. Hui, R. Johnson, T. S. Lo, Phys. Rev. Lett. {\bf 82}, 3360 (1999).


\bibitem{lo1} T. S. Lo, S. W. Lim, P. M. Hui, N. F. Johnson, Physica A {\bf 287}, 313 (2000).

\bibitem{lo2} T. S. Lo, P. M. Hui, N. F. Johnson, Phys. Rev. E {\bf 62}, 4393 (2000).

\bibitem{hod1} S. Hod, E. Nakar, Phys. Rev. Lett. {\bf 88}, 238702 (2002)

\bibitem{hod2} S. Hod, E. Nakar, ``Temporal oscillations and phase transitions in the evolutionary minority game'', arXiv: cond-mat/020656

\bibitem{hod3} S. Hod, ``Time-dependent random walks and the theory of complex adaptive systems", arXiv: cond-mat/0212055

\bibitem{burgos} E. Burgos, H. Ceva, R. P. J. Perazz, Comment on ``Self-segregation versus clustering in the evolutionary minority game'', arXiv:cond-mat/0301518

\bibitem{hod4} S. Hod, E. Nakar, ``Strategy updating rules and strategy distributions in dynamical multiagent systems", arXiv: cond-mat/032099


\end{thebibliography}
\end{document}